\documentclass[twocolumn]{aastex701}

\begin{document}

\title{The Impact of Population III.1 Flash Reionization for\\ CMB Polarization and Thomson Scattering Optical Depth}

\author[orcid=0000-0002-3389-9142]{Jonathan C. Tan}
\affiliation{Dept. of Physics \& Astronomy, Chalmers University of Technology, Gothenburg, Sweden}
\affiliation{Dept. of Astronomy \& Virginia Institute for Theoretical Astronomy, University of Virginia, Charlottesville, VA, USA}
\email{jctan.astro@gmail.com}  

\author[orcid=0000-0002-0136-2404]{Eiichiro Komatsu}
\affiliation{
Max-Planck-Institut f\"ur Astrophysik, Karl-Schwarzschild-Str. 1, 85741 Garching, Germany
}
\affiliation{
Ludwig-Maximilians-Universit\"at M\"unchen, Schellingstr. 4, 80799 M\"unchen, Germany
}
\affiliation{
Kavli Institute for the Physics and Mathematics of the Universe (WPI), The University of Tokyo Institutes for Advanced Study (UTIAS), The University of Tokyo, Chiba 277-8583, Japan
}
\email{komatsu@MPA-Garching.MPG.DE}

%

\begin{abstract}
The Population III.1 theory for supermassive black hole (SMBH) formation predicts a very early ($z\sim20-25$) transient phase, the ``Pop III.1 Flash'', of cosmic reionization powered by supermassive stars that are SMBH progenitors. The universe then quickly recombined to become mostly neutral, with this state persisting until galaxies begin to reionize intergalactic gas again at $z\sim 10$. The overall Thomson scattering optical depth, $\tau$, from the Pop III.1 Flash has been shown to be $\tau_{\rm PopIII.1}\sim0.03$, leading to a total $\tau\sim0.08-0.09$. Such a value, while significantly larger than that previously inferred from {\it Planck} observations of the low-$l$ $EE$ polarization power spectrum of the CMB, can help relieve several ``tensions'' faced by the standard $\Lambda$CDM cosmological model, especially the preference for negative neutrino masses and dynamic dark energy. Here we compute $EE$ power spectra of example models of the Pop III.1 Flash. We find that, because of its very high redshift, the contribution to $l\lesssim\:$6 modes is dramatically reduced compared to usual low-$z$ reionization models for the same value of $\tau$, while the power at $l\gtrsim\:$6 is boosted. Thus the Pop III.1 reionization scenario provides a natural way to increase $\tau$, while remaining closer to the latest CMB low-$l$ polarization observations.
\end{abstract}



\section{Introduction}\label{sec:intro}

The epoch of cosmic reionization -- the first billion years after the Big Bang -- is a frontier of astrophysical and cosmological research today. Intense ultraviolet radiation from the first ``Population III'' stars, which were likely formed out of pristine hydrogen and helium gas in pre-galactic dark matter halos, reionized the surrounding neutral atoms \citep[see][for a review]{2001PhR...349..125B}. However, we know very little about the time, process, and consequences of the formation of these first stellar sources. 

The Population III.1 (hereafter Pop III.1) theory for supermassive black hole (SMBH) formation \citep{2019MNRAS.483.3592B,2023MNRAS.525..969S,2025MNRAS.536..851C} \citep[see also][for a review]{2024arXiv241201828T} proposes that SMBH-progenitor supermassive, i.e., $\sim10^5\:M_\odot$, stars are born in ``Pop III.1'' minihalos, i.e., the first-forming, locally-isolated, metal-free dark matter halos with masses $\sim10^6\:M_\odot$ that are undisturbed by external feedback. These characteristics allow for formation of a single massive star co-located with a highly concentrated dark matter cusp. Then the process of weakly interacting massive particle (WIMP) dark matter annihilation in the Pop III.1 protostar can affect its structure \citep{2008PhRvL.100e1101S,2009ApJ...692..574N}, in particular giving it a large photospheric radius with a relatively cool temperature \citep{2015ApJ...799..210R,2025arXiv250700870N,2025arXiv251000216T}. This allows the growing star to avoid early photoevaporative ionizing feedback that for ``standard'' Pop III stars, i.e., those not impacted by WIMP annihilation, typically truncates accretion around $100\:M_\odot$ during contraction to the zero age main sequence \citep{2008ApJ...681..771M,2010AIPC.1294...34T,2011Sci...334.1250H,2014ApJ...792...32S,2014ApJ...781...60H}. Thus Pop III.1 protostars are theorized to be able to grow to $\sim 10^5\:M_\odot$, with this mass scale set by the baryonic content of the dark matter minihalo. 

In the final stages of their evolution before they collapse to a SMBH, Pop III.1 supermassive stars are expected to become powerful sources of H-ionizing photons. These propagate quickly through the intergalactic medium (IGM), creating ionized regions ({\it R-type} HII regions), with typical radii of $\sim 1$ comoving megaparsecs (cMpc). The feedback from these HII regions then sets the cosmic abundance of  SMBHs, since Pop III.1 sources are required to have formed in pristine gas that was not impacted by a high level of ionization. In the fiducial case, the overall abundance of SMBHs is expected to be $n_{\rm SMBH}\sim10^{-1}\:{\rm cMpc}^{-3}$ \citep{2024arXiv241201828T}. The phase of Pop III.1 star formation ends by $z\sim20$ (about 200~Myr after the Big Bang), after which the IGM quickly recombines. The gas only begins to be substantially ionized again once sufficient luminous galaxies have formed by $z\sim10$ (about 500~Myr after the Big Bang).

The Pop III.1 theory thus predicts an early phase of ionization of the universe, referred to as the ``Pop III.1 Flash'', in which a large fraction of the IGM at redshifts $z\sim20-25$ was impacted.
The contribution of the Pop III.1 Flash to the overall Thomson scattering optical depth of observed CMB photons has been evaluated to be $\tau_{\rm PopIII.1}\sim 0.03$ \citep{2025ApJ...989L..47T}. This is a separate contribution from that associated with reionization of the universe at lower redshifts ($z\lesssim10$) by normal galaxies and active galactic nuclei (AGN) \citep{2015ApJ...802L..19R,2017MNRAS.465.4838G,2024ApJ...975..208T,2024MNRAS.535L..37M,2025ApJS..278...33K,2025ApJ...981..134P,2025MNRAS.542.1952M}. The ionized IGM from this late-phase reionization is known to produce an optical depth of $\tau_{\rm gal}\simeq 0.06$. Such a level is consistent with previous results derived from the amplitude of the low-$l$ 
$EE$-modes of the CMB from the Planck Collaboration \citep{2020A&A...641A...6P}, who found $\tau=0.054\pm0.007$. Subsequent reanalyses of \textit{Planck} data yielded slightly larger values, $\tau=0.058\pm 0.006$ \citep{Tristram:2023haj} and $0.0063\pm 0.005$ \citep{2021MNRAS.507.1072D}. Parametric reionization histories that are constrained by these CMB data indicate that this late-phase reionization began at $z\sim10$ and ended at $z\sim 5$.

While the late-phase reionization by standard galaxies is consistent with the latest CMB results with $\tau\simeq0.06$, a number of recent papers, which did not rely on the low-$l$ polarization data, have argued for a larger value of $\tau\simeq 0.09$  \citep{2025JCAP...08..082A,2025PhRvD.112d3541J,2026PhRvL.136h1002S,2026PhRvD.113h3515A}. Such higher values would help alleviate tensions arising from CMB-based estimates of the Hubble constant (i.e., ``Hubble tension'') and from recent Baryonic Acoustic Oscillation measurements from the Dark Energy Spectroscopic Instrument (DESI) \citep{2025arXiv250314738D}, 
which, combined with {\it Planck} CMB results, manifest as a preference for negative neutrino masses (or positive, but too small masses)  and evolving, i.e., dynamic, dark energy. It has been pointed out by the above authors that the measurement of $\tau$ from the CMB faces a number of challenging systematic uncertainties, especially instrumental systematic effects \citep{Planck:2013dkx,Planck:2015aiq,Planck:2016kqe}, which might yet allow compatibility with a larger value. Nevertheless, as we discuss below, values of $\tau\simeq0.09$ with standard late-phase reionization histories produce amplitudes of low-$l$ $EE$ modes that appear difficult to reconcile with {\it Planck}.

In this {\it Letter} we compute $EE$ power spectra of example models of Pop III.1 reionization, i.e., with a very early phase of flash ionization, combined with standard late-phase contributions. We find that, because of the very high redshift of the Pop III.1 ionization phase, the impact on the $l\lesssim6$ CMB $EE$ modes is dramatically reduced compared to the usual low-$z$ contribution for the same value of $\tau$, while the power at $l\gtrsim6$ is boosted. We describe our methods in \S\ref{sec:methods}, our results in \S\ref{sec:results}, and discuss their implications in \S\ref{sec:discussion}.

\section{\label{sec:methods}Methods}

We parameterize the reionization history of the universe with two components. For the late-phase, ``low-$z$'' component we adopt a standard ${\rm tanh}(z)$ function to describe the ionization fraction (defined as the ratio of the number density of free electrons to that of H nuclei) \citep{Lewis:2008wr}:
\begin{equation}
x_e(y) = \frac{f}{2}\left[1 + {\rm tanh}\left(\frac{y(z_{\rm re})-y}{\Delta_y} \right)  \right],
\label{eq:rion}
\end{equation}
where $y(z_{\rm re})=(1+z_{\rm re})^{3/2}$ is the location where $x_e=f/2$, i.e., the ``midway'' point toward the maximum level of reionization, $\Delta_y = 1.5\sqrt{1+z_{\rm re}}\Delta_z$, where $\Delta_z$ parameterizes the width of the transition, and, for the case of singly and doubly ionized He, $f=1+(n_{\rm He}/n_{\rm H})\simeq 1.08$ and $1+(2n_{\rm He}/n_{\rm H})\simeq1.16$, respectively. For low-$z$ reionization we adopt a standard treatment of He reionization: He is singly reionized when H is reionized and doubly reionized at $z = 3.5$ \citep{Lewis:2008wr}.
We will also use the metric $x_{i,{\rm H}}$, which is the ionization fraction of H described by Eq.~\ref{eq:rion} with $f=1$.

For the high-$z$ reionization associated with the Pop III.1 Flash, we follow the model of \cite{2025ApJ...989L..47T}. This assumes supermassive Pop III.1 stars form at a redshift $z_{\rm form}$, leading to a subsequent peak level of ionization in their HII regions, $x_{i,{\rm H,peak}}$, at redshift $z_{\rm flash}$. These HII regions fill a significant volume fraction of the IGM, $f_{i,{\rm vol}}$. Following the results of the cosmological Pop III.1 SMBH seeding models \citep{2019MNRAS.483.3592B,2023MNRAS.525..969S}, we consider two cases: $z_{\rm flash}=20$ and 25. Note that these values are constrained by the requirement of forming sufficient SMBHs to match the observed population with $n_{\rm SMBH}\sim 10^{-2}-10^{-1}\:{\rm cMpc}^{-3}$ \citep{2024ApJ...971L..16H,2025ApJ...991..141C}.

For a fiducial case we consider values of $x_{i,{\rm H,peak}}=1.0$ and $f_{i,{\rm vol}}=0.5$, noting that there is a simple linear relationship between these parameters for the total contribution to $\tau$. However, in the context of the Pop III.1 model, we require $f_{i,{\rm vol}}$ to be near unity. Similarly, HII regions undergoing {\it R-type} evolution are expected to have ionization fractions of order unity when the star is shining \citep{2025MNRAS.542.1532S}. Our value of $x_{i,{\rm H,peak}}$ is a factor of two higher than that adopted by \cite{2025ApJ...989L..47T}, which is motivated by our more accurate calculation of $\tau$ (see below).

The timescale for the ionization fraction to rise up to its peak value, $t_{\rm rise}$, is assumed to lie between the lifetime of the ionizing source and the time to establish a Strömgren sphere, $t_{\rm ion}$, in gas that has a density similar to that of the mean IGM density. For the former, we take a fiducial value of 10~Myr for the lifetime of a supermassive star that is somewhat prolonged by WIMP annihilation heating. For the latter, the value of $t_{\rm ion}$ assuming the mean cosmic density is
\begin{equation}
    t_{\rm ion} =\frac43 \pi R_S^3 \frac{n_{\rm H}}{S}=\frac{1}{\alpha^{(2)}n_{\rm H}}\simeq 51.3\left(\frac{1+z_{\rm form}}{31}\right)^{-3}\:{\rm Myr},\label{eq:tion}
\end{equation}
where $R_S$ is the radius of the Strömgren sphere, $n_{\rm H}$ is the total number density of H nuclei, $\alpha^{(2)}=1.08\times 10^{-13}(T/30,000\:{\rm K})^{-0.8}\:{\rm cm^3\:s}^{-1}$ is the recombination rate to excited states of ionized H, where we have normalized to a fiducial temperature of $T=30,000\:$K expected in metal-free gas, and $S$ is the rate of production of H-ionizing photons.
The final calculation adopts the mean number density of H nuclei at $z=30$, i.e., $n_{\rm H,z=30}=5.72\times10^{-3}\:{\rm cm}^{-3}$. For simplicity and given the uncertainty in Pop III.1 star ionizing spectra, we have ignored the reionization of He in this high-$z$ reionization phase, although it is included in the low-$z$ reionization component described by Eq.~\ref{eq:rion}.

Pop III.1 sources are expected to form in moderately overdense regions. For example, in the simulations of \cite{2025MNRAS.542.1532S} the average density in the HII region around a Pop III.1 supermassive star is a factor of about three times greater than the cosmic average. From such simulations of ionizing feedback it is also seen that when $t_{\rm ion}>t_*$, the {\it R-type} ionization front continues to propagate after the star has ended its life. Given the above considerations we adopt a fiducial value for $t_{\rm rise}=30\:$Myr. This implies that Pop III.1 stars would actually have started shining at $z_{\rm form}\simeq 23$ and 30 for our cases of $z_{\rm flash}=20$ and 25, respectively. Thus $t_{\rm rise}$ can additionally be interpreted as encompassing a modest spread in formation redshifts of Pop III.1 stars that is similar in magnitude to these differences.

After reaching a peak ionization fraction in the Pop III.1 Flash, we assume that this level decreases exponentially on a timescale equal to the recombination timescale in the HII regions given by Eq.~\ref{eq:tion}, but adopting an overdensity compared to the mean IGM of a factor of three. Note, given the {\it R-type} nature of the HII regions, there is limited impact on the density structure of the gas due to the ionizing feedback. For our cases of $z_{\rm flash}=20$ and 25, these recombination times are 55 and 29~Myr, respectively.

We use a linear Boltzmann solver \texttt{CLASS} \citep{Blas:2011rf} to calculate the CMB $EE$ power spectra for the various reionization histories, along with their associated values of $\tau$.
We assume a flat $\Lambda$ cold dark matter ($\Lambda$CDM) cosmological model with its parameters taken from `Plik best fit' in Table 1 of \citep{Planck:2020olo}: ($h$, $\Omega_{\rm c}h^2$, $\Omega_{\rm b}h^2$, $\ln(10^{10}A_{\rm s})$, $n_{\rm s}$) = (0.6732, 0.12011, 0.022383, 3.0448, 0.96605), with the minimal total mass for neutrinos (0.06~eV). The total mass density parameter including the neutrino is $\Omega_{\rm m}=0.3158$ and the cosmological constant is given by $\Omega_\Lambda = 1 - \Omega_{\rm m}$.

\section{\label{sec:results}Results}

\begin{figure}
\includegraphics[width=1.0\linewidth]{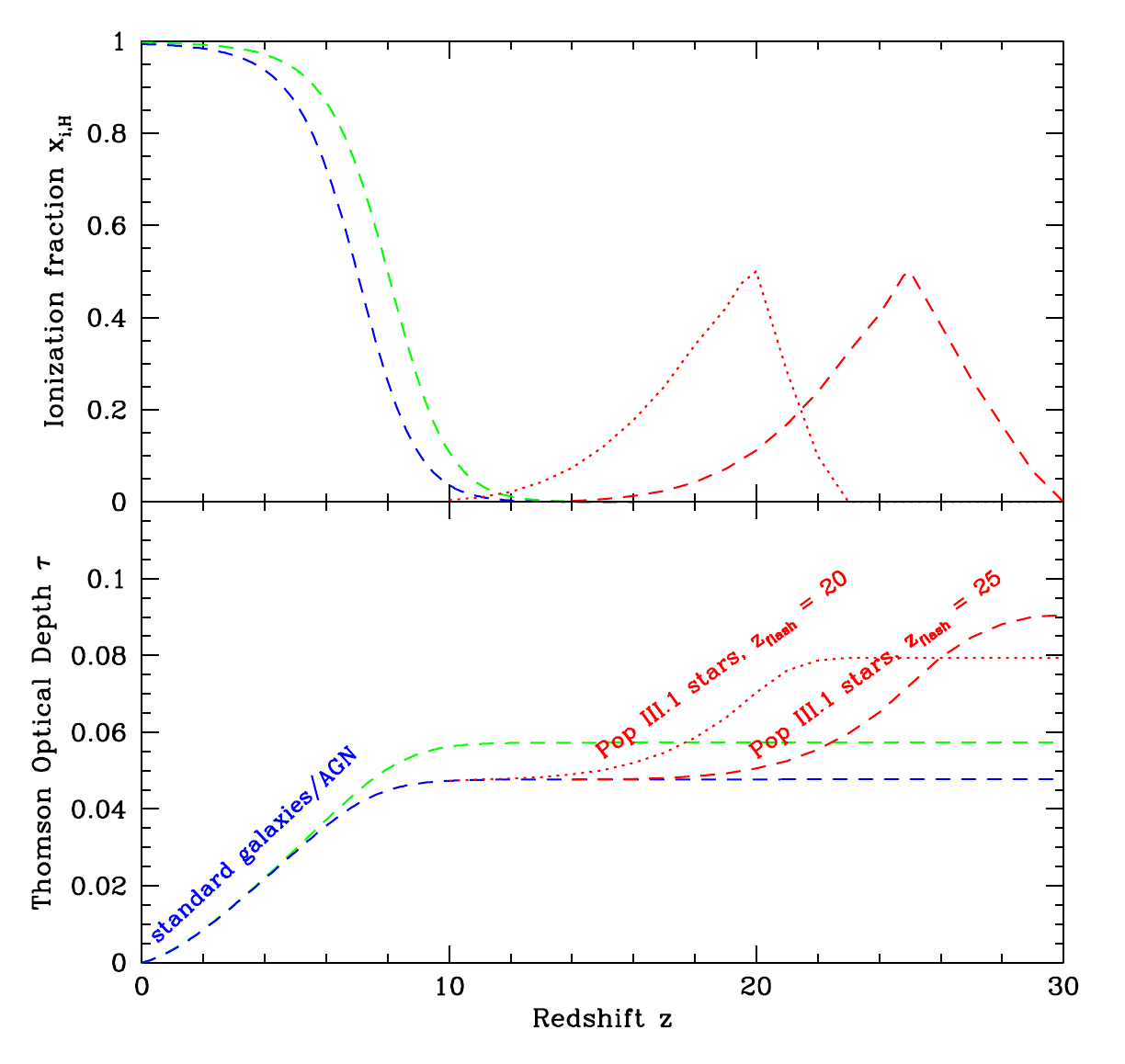}
\caption{\label{fig:tau} {\it (a) Top:} Ionization fraction history of the universe, shown as a function of redshift, with $x_{i,{\rm H}}$ measuring the fraction of H atoms that are ionized. The green dashed line shows a standard reionization history described with a ${\rm tanh} (z)$ function with $z_{\rm re}=8$ [Eq.~\ref{eq:rion}], from fully ionized conditions at low redshift to fully neutral conditions at high redshift. 
The dashed blue line is an equivalent model, but with $z_{\rm re}=7$. Note, these models include a standard treatment of He reionization (see text).
The red lines show fiducial ionization histories resulting from Pop III.1 supermassive stars, peaking at $z_{\rm flash}=20$ (dotted) and 25 (dashed). {\it (b) Bottom:} Thomson optical depth to electron scattering, $\tau$, integrated out to redshift, $z$. The green and blue dashed lines show the ${\rm tanh}(z)$ models with $z_{\rm re}=8$ and 7, respectively, representing the contribution from ``standard'' galaxies and AGN. The red dotted and dashed lines show the contribution from Pop III.1 supermassive stars with epoch of peak flash ionization at $z_{\rm flash}=20$ and 25, respectively, which have been combined with the ${\rm tanh} (z)$ with $z_{\rm re}=7$ model.} 
\end{figure}

\begin{figure}
\includegraphics[width=1.0\linewidth]{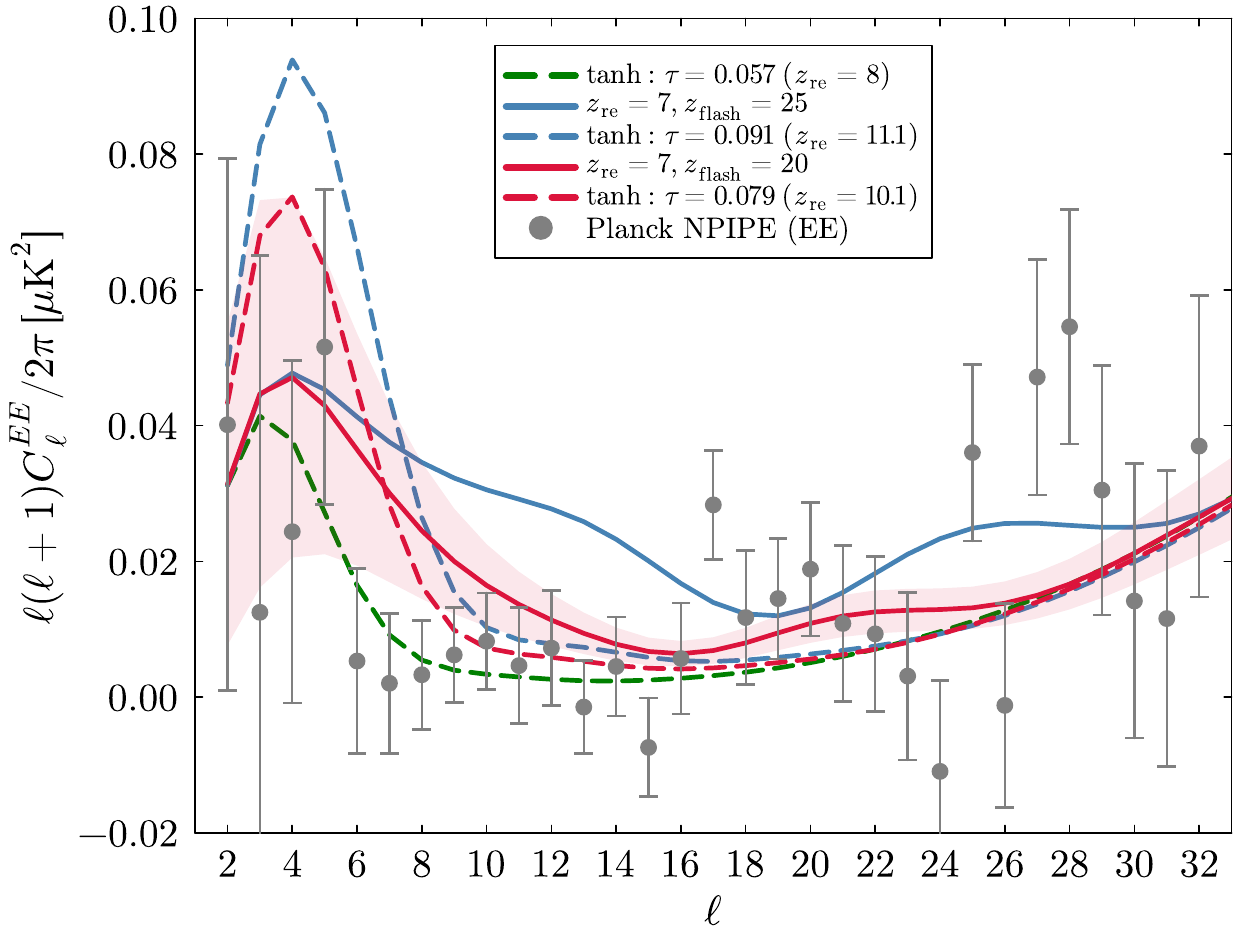}
\caption{\label{fig:cmb} CMB $EE$ power spectra for various reionization models. The green dashed line shows the ${\rm tanh}(z)$ model with $z_{\rm re}=8$, yielding a total $\tau=0.057$. The red and blue dashed lines show equivalent ${\rm tanh}(z)$ models with $\tau$ of 0.079 and 0.091, respectively. The red and blue solid lines show fiducial ${\rm tanh}(z)$ + Pop III.1 reionization models with $z_{\rm flash}=20$ and 25 that also yield $\tau$ of 0.079 and 0.091, respectively. The shaded region around the red solid line shows a cosmic-variance-limited error range to the $z_{\rm flash}=20$ model, assuming 70\% sky coverage, i.e., a reasonable approximation for expected  LiteBIRD uncertainties. The grey circles and error bars show the {\it Planck} NPIPE data \citep{Planck:2020olo,Tristram:2020wbi}.
}
\end{figure}

Figure~\ref{fig:tau} shows some example reionization histories and associated cumulative distributions of $\tau(z)=c\sigma_Tn_{\rm H0}\int_0^z \mathrm{d}z~ x_e(z)(1+z)^2H^{-1}(z)$, where $c$ is the speed of light, $\sigma_T$ the Thomson scattering cross section, $n_{\rm H0}$ the present-day number density of protons, and $H(z)$ the Hubble expansion rate.
For the cases of late-phase reionization by ``standard galaxies and AGN'', these ``${\rm tanh}(z)$'' models are parameterized by $z_{\rm re}$ and $\Delta_z=2$ [Eq.~\ref{eq:rion}]. 
The green and blue dashed lines show models with $z_{\rm re}=8$ and 7, which yield total values of $\tau=0.0573$ and 0.0477, respectively.
To the latter model we also add on the high-$z$ Pop III.1 Flash component with $z_{\rm flash}=20$ and 25 (both with $x_{i,{\rm H,peak}}=1.0$ and $f_{i,{\rm vol}}=0.5$), which yield total values of $\tau=0.0794$ and 0.0906, respectively.

Figure~\ref{fig:cmb} shows the low-$l$ $EE$ CMB power spectra from various reionization models compared to the {\it Planck} Public Release 4 (PR4) ``NPIPE'' data \citep{Planck:2020olo,Tristram:2020wbi}. The green dashed line shows the ${\rm tanh}(z)$ model with $z_{\rm re}=8$, which yields a good fit to the {\it Planck} data. 

The red lines show models with $\tau$ elevated to 0.079. The dashed line shows the $EE$ power spectrum when this $\tau$ is achieved via a standard ${\rm tanh}(z)$ function (requiring $z_{\rm re}=10.1$). The solid line shows a model with ${\rm tanh}(z)$ with $z_{\rm re}=7$ combined with a fiducial Pop III.1 contribution with $z_{\rm flash}=20$. We find that even though these models have the same total value of $\tau$, the amplitudes of the lowest $l$ modes $(l\lesssim6)$ are significantly smaller for the model with the Pop III.1 component. Focusing on these $l\leq6$ modes, i.e., those most commonly used to constrain cosmic reionization, then the Pop III.1 Flash model has $\chi^2 = 6.7$, while the ${\rm tanh}(z)$-only model has $\chi^2 = 13.6$.

Similarly, the blue lines show two models with $\tau=0.091$. The dashed line shows a case with a ${\rm tanh}(z)$ function (requiring $z_{\rm re}=11.1$). The solid line shows the case with ${\rm tanh}(z)$ with $z_{\rm re}=7$ combined with a fiducial Pop III.1 model with $z_{\rm flash}=25$. The model with the Pop III.1 flash component again has much lower power at $l\lesssim6$. However we note that at intermediate scales with $7\leq l\leq18$ this model has significantly higher amplitudes that appear to be in tension with the latest {\it Planck} data.

This behavior can be explained as follows. CMB polarization is generated when an electron at a redshift of $z$ scatters the quadrupole of the CMB as seen by the electron. The comoving wavenumber of the quadrupole as seen by the electron is given by $k\simeq 3/[r_L - r(z)]$, where $r_L = 14~\mathrm{cGpc}$ is the comoving distance from Earth to
our last-scattering surface, and $r(z)$ is the comoving distance to redshift $z$. For example, a redshift of $z = 8$ yields $r(8) \simeq 9$~cGpc. We observe this wavenumber at a multipole of $l \le  kr(8)\simeq 6$, which corresponds to the so-called ``reionization bump'' in the polarization power spectra \citep{Zaldarriaga:1996ke}. As $r(z)$ increases with $z$, contributions to the $EE$ power spectrum spread to larger values of $l$ for sources at larger $z$. For a redshift of $z=20$ relevant to the Pop III.1 Flash, we have $r(20)\simeq 11\:$cGpc. Thus, $k\simeq 1~\mathrm{cGpc}^{-1}$ and the multipole is observed at $l \le  kr(20)\simeq 11$. We note that the secondary ``peaks'' at $l\sim22$ and 26 for $z_{\rm flash}=20$ and 25 are the result of an oscillation due to a spherical Bessel function of order two.

\section{\label{sec:discussion}Discussion and Conclusions}

We have shown that reionization models including a component of Pop III.1 flash ionization at redshifts of 20 to 25 yield CMB $EE$-mode power spectra with very different shapes compared to equivalent-$\tau$ standard late-phase reionization models described with a ${\rm tanh}(z)$ function [Eq.~\ref{eq:rion}]. In particular, if a certain elevated value of $\tau\simeq 0.08-0.09$ is desired, e.g., to alleviate tensions of ``negative neutrino masses'' or ``dynamic dark energy'' \citep{2025JCAP...08..082A,2025PhRvD.112d3541J,2026PhRvL.136h1002S,2026PhRvD.113h3515A}, then Pop III.1 models offer a way to do this that minimizes tension with the observed low-$l$ CMB $EE$ mode constraints.

Sensitivity of the CMB $EE$ power spectrum to the redshift dependence of reionization history has been discussed previously \citep[e.g.,][]{Zaldarriaga:2008ap,2017PhRvD..95b3513H,2018A&A...617A..96M}. However, the particular scenario of very early Pop III.1 flash ionization has not been discussed in detail.

Future CMB polarization observations, e.g., with CLASS \citep{2025ApJ...986..111L} and LiteBIRD \citep{2023PTEP.2023d2F01L}, are expected to be able to improve on the low-$l$ $E$-mode constraints and thus test the proposed Pop III.1 reionization history. As we see from Figure~\ref{fig:cmb}, achieving an accuracy of $\sim20\%$ in the range $10\lesssim l \lesssim 30$ would be sufficient to distinguish different Pop III.1 models with $z_{\rm flash}$ in the range 20 to 25. LiteBIRD will deliver measurements of the $EE$ power spectrum with a precision limited only by cosmic variance up to $l\simeq 200$. Such data will easily distinguish between the models shown in Figure~\ref{fig:cmb}.

Future studies of the CMB with the Simons Observatory \citep{2019JCAP...02..056A} are expected to be able to further test the Pop III.1 prediction of an early phase of flash ionization, especially via observations of the patchy kinematic Sunyeav-Zel'dovich (pkSZ) effect from the peculiar motion of the HII regions. We note that current pkSZ constraints on reionization history, which have been found to be in $2\sigma$ tension with values of $\tau\simeq0.09$ \citep{2025ApJ...987L..29C}, have assumed a monotonically decreasing IGM ionization fraction with redshift. These need to be re-calculated for the Pop III.1 scenario with a distinct phase of very early flash ionization. We also note that the peculiar motions driving the pkSZ signal are reduced at higher redshifts, so the Pop III.1 reionization scenario has an attractive feature of being able to boost $\tau$ in a way that minimizes additional pkSZ contributions.
Furthermore, the fiducial Pop III.1 model predicts the formation of ionized bubbles with a typical size of 1 cMpc at $z\simeq 20-25$. These bubbles may leave an imprint on the pkSZ power spectrum \citep{Park:2013mv}.

There are several other implications of the Pop III.1 model of early reionization, which have also been discussed by \cite{2025ApJ...989L..47T}. These include signatures in high-$z$ 21-cm emission that arise from ionized bubbles and/or heating effects on neutral gas around Pop III.1 sources. There is also a potential impact on the strength of the sky-averaged (global) 21-cm signal of neutral hydrogen absorption from the early universe. A tentative detection of this signal, centered at $z=17.2$, has been claimed based on analysis of data from the low-band antenna of the Experiment to Detect the Global EoR Signature (EDGES) \citep{2018Natur.555...67B} \citep[however, see][]{2022NatAs...6..607S}. However, the absorption depth seen is at least twice as strong as predicted by standard astrophysical scenarios in $\Lambda$CDM. A potential resolution of this anomalous absorption is an enhanced radio background, equivalent to a brightness temperature of 67.2~K, significantly greater than that of the CMB at these redshifts with $T_{\rm CMB}=49.5\:$K \citep[e.g.,][]{2018ApJ...858L..17F,2019MNRAS.486.1763F}.
\cite{2025ApJ...989L..47T} has presented a simple estimate of the expected free-free emission from the Pop III.1 Flash, finding that for reasonable choices of parameters, it is able to produce a radio background strong enough to explain the deep absorption depth of the EDGES result.

Finally, the most fundamental implication of the Pop~III.1 cosmological model for SMBH formation is that it requires energy input from self-annihilating dark matter in the very first generation of stars. In the framework of $\Lambda$CDM cosmology, the most natural way for this to occur is via capture and decay of WIMPs in primordial protostars \citep{2008PhRvL.100e1101S,2009ApJ...692..574N,2015ApJ...799..210R,2025arXiv250700870N,2025arXiv251000216T}, with particle masses of $m_\chi\sim100-1,000\:$GeV and self-annihilation rate coefficients of $\langle \sigma_a v \rangle \simeq
3 \times 10^{-26}\:{\rm cm^3\:s}^{-1}$ \citep[e.g.,][]{1996PhR...267..195J} to explain the cosmic dark matter density, $\Omega_{\rm DM} = 0.268$. Thus a basic prediction of the Pop III.1 model (and more generally of such $\Lambda$CDM models) is the existence of WIMPs with the above properties, with implications for direct and indirect dark matter search experiments. In this context, the recent claim \citep{2025JCAP...11..080T} of a $\sim20\:$GeV gamma-ray excess seen in Fermi Large Area Telescope (LAT) data in the halo region of the Milky Way, which implies a WIMP mass of $\sim 500\:$GeV, is intriguing and warrants further investigation.
 
\begin{acknowledgments}
We thank Richard Ellis for helpful discussions and Matthieu Tristram for his help with the \textit{Planck} NPIPE data. We also thank ``Giesinger Br\"au Stehausschank'' for their hospitality, where the original idea for this project emerged. We thank the referee for helpful and constructive comments. We thank Robert Lupton and his collaborators for developing the SuperMongo graphics software and Ned Wright for developing his Java Script Cosmology Calculator \citep{2006PASP..118.1711W}. J.C.T. acknowledges funding from the Chalmers Initiative on Cosmic Origins (CICO), the Virginia Initiative on Cosmic Origins (VICO), and the Virginia Institute for Theoretical Astrophysics (VITA), supported by the College and Graduate School of Arts and Sciences at the University of Virginia. E.K.'s work was supported in part by the Excellence Cluster ORIGINS which is funded by the Deutsche Forschungsgemeinschaft (DFG, German Research Foundation) under Germany’s Excellence Strategy: Grant No.~EXC-2094 - 390783311. The Kavli IPMU is supported by World Premier International Research Center Initiative (WPI), MEXT, Japan.
\end{acknowledgments}

\bibliography{PopIII.1_cmb_apjl}{}
\bibliographystyle{aasjournalv7}



\end{document}